\documentclass[sigconf,natbib=true, screen=true]{acmart}

\acmSubmissionID{9455}

\usepackage{colortbl}
\usepackage{arydshln}
\usepackage{multirow}
\usepackage{multicol}
 \usepackage{array}
\usepackage{bm}
\usepackage{booktabs} 
\usepackage{amsmath}
\usepackage{amsfonts}
\usepackage{graphicx}
\usepackage{tabularx}
\usepackage{textcomp}
\usepackage{xcolor}
\usepackage{listings}
\usepackage{multirow}
\usepackage{subfig}
\usepackage{graphicx}
\usepackage{bbding}
\usepackage[inline]{enumitem}
\usepackage[ruled,linesnumbered]{algorithm2e}
\usepackage[skip=0pt]{caption}
\usepackage{rotating}
\usepackage[para]{footmisc}
\usepackage{xspace}

\AtBeginDocument{%
  \providecommand\BibTeX{{%
    \normalfont B\kern-0.5em{\scshape i\kern-0.25em b}\kern-0.8em\TeX}}}

\newcommand{\ie}{i.e.,}
\newcommand{\eg}{e.g.,}

\newcommand{\ours}{ToolRec\xspace}
\newcommand{\wrt}{w.r.t.\xspace}
\newcommand{\cf}{cf.\xspace}

\newcommand{\header}[1]{\vspace*{1mm}\noindent\textbf{#1.}}

\newcommand{\negskip}{\vspace*{-1.5mm}}

\copyrightyear{2024}
\acmYear{2024}
\setcopyright{acmlicensed}
\acmConference[SIGIR '24]{Proceedings of the 47th International ACM SIGIR Conference on Research and Development in Information Retrieval}{July 14--18, 2024}{Washington, DC, USA}
\acmBooktitle{Proceedings of the 47th International ACM SIGIR Conference on Research and Development in Information Retrieval (SIGIR '24), July 14--18, 2024, Washington, DC, USA}
\acmDOI{10.1145/3626772.3657828}
\acmISBN{979-8-4007-0431-4/24/07}



\begin{document}

\title{Let Me Do It For You: Towards LLM Empowered Recommendation via Tool Learning}


\author{Yuyue Zhao}
\orcid{0000-0002-5298-0309}
\email{yyzha0@mail.ustc.edu.cn}
\affiliation{%
  \institution{University of Science and Technology of China}
  \institution{University of Amsterdam}
\city{Hefei}
\state{Anhui}
  \country{China}
  }

\author{Jiancan Wu}
\orcid{0000-0002-6941-5218}
\authornote{Jiancan Wu and Xiang Wang are corresponding authors.}
\email{wujcan@gmail.com}
\affiliation{%
  \institution{University of Science and Technology of China}
  \city{Hefei}
\state{Anhui}
  \country{China}
  }

\author{Xiang Wang}
\authornotemark[1]
\orcid{0000-0002-6148-6329}
\authornote{Xiang Wang is also affiliated with Institute of Dataspace, Hefei Comprehensive National Science Center.}
\email{xiangwang1223@gmail.com}
\affiliation{%
  \institution{University of Science and Technology of China}
  \city{Hefei}
\state{Anhui}
  \country{China}
  }
  


\author{Wei Tang}
\orcid{0000-0001-6561-7026}
\email{weitang@mail.ustc.edu.cn}
\affiliation{%
  \institution{University of Science and Technology of China}
    \city{Hefei}
\state{Anhui}
  \country{China}
  }

\author{Dingxian Wang}
\orcid{0000-0002-6880-7869}
\email{dingxian.wang@student.uts.edu.au}
\affiliation{%
  \institution{University of Technology Sydney}
    \city{Ultimo}
\state{New South Wales}
  \country{Australia}
  }

\author{Maarten de Rijke}
\orcid{0000-0002-1086-0202}
\email{m.derijke@uva.nl}
\affiliation{%
  \institution{University of Amsterdam}
  \city{Amsterdam}
  \country{The Netherlands}
  }

\def\authors{Yuyue Zhao, Jiancan Wu, Xiang Wang, Wei Tang, Dingxian Wang, and\linebreak Maarten de Rijke}

\renewcommand{\shortauthors}{Yuyue Zhao et al.}

\begin{abstract}

Conventional recommender systems (RSs) face challenges in precisely capturing users' fine-grained preferences.
Large language models (LLMs) have shown capabilities in commonsense reasoning and leveraging external tools that may help address these challenges. However, existing LLM-based RSs suffer from hallucinations, misalignment between the semantic space of items and the behavior space of users, or overly simplistic control strategies (\eg\ whether to rank or directly present existing results). To bridge these gap, we introduce \ours, a framework for LLM-empowered recommendations via tool learning that uses LLMs as surrogate users, thereby guiding the recommendation process and invoking external tools to generate a recommendation list that aligns closely with users' nuanced preferences.

We formulate the recommendation process as a process aimed at exploring user interests in attribute granularity. 
The process factors in the nuances of the context and user preferences. 
The LLM then invokes external tools based on a user's attribute instructions and probes different segments of the item pool. 
We consider two types of attribute-oriented tools: rank tools and retrieval tools. 
Through the integration of LLMs, \ours enables conventional recommender systems to become external tools with a natural language interface.
Extensive experiments verify the effectiveness of \ours, particularly in scenarios that are rich in semantic content. 
\end{abstract}

\maketitle

\section{Introduction} \label{sec:intro}

Recommender systems (RSs) are typically designed to identify user preferences and subsequently suggest potential items of interest~\cite{rendle2012bpr, sasrec, he2020lightgcn, wang2019kgat, WuWF0CLX21, Wu2024SSM, YangWWWY023, GaoWLCHLLZJ24}.
This strategy has two important limitations. 
First, the capabilities of existing RSs to accurately capture a user's true preferences are limited when relying solely on historical interaction data. 
Second, conventional RSs are often ``narrow experts,'' lacking commonsense knowledge about users and items, which leads to a restricted scope of recommendations~\cite{kgrecsurvey}.

\begin{figure}[tb]
\includegraphics[width=1\columnwidth]{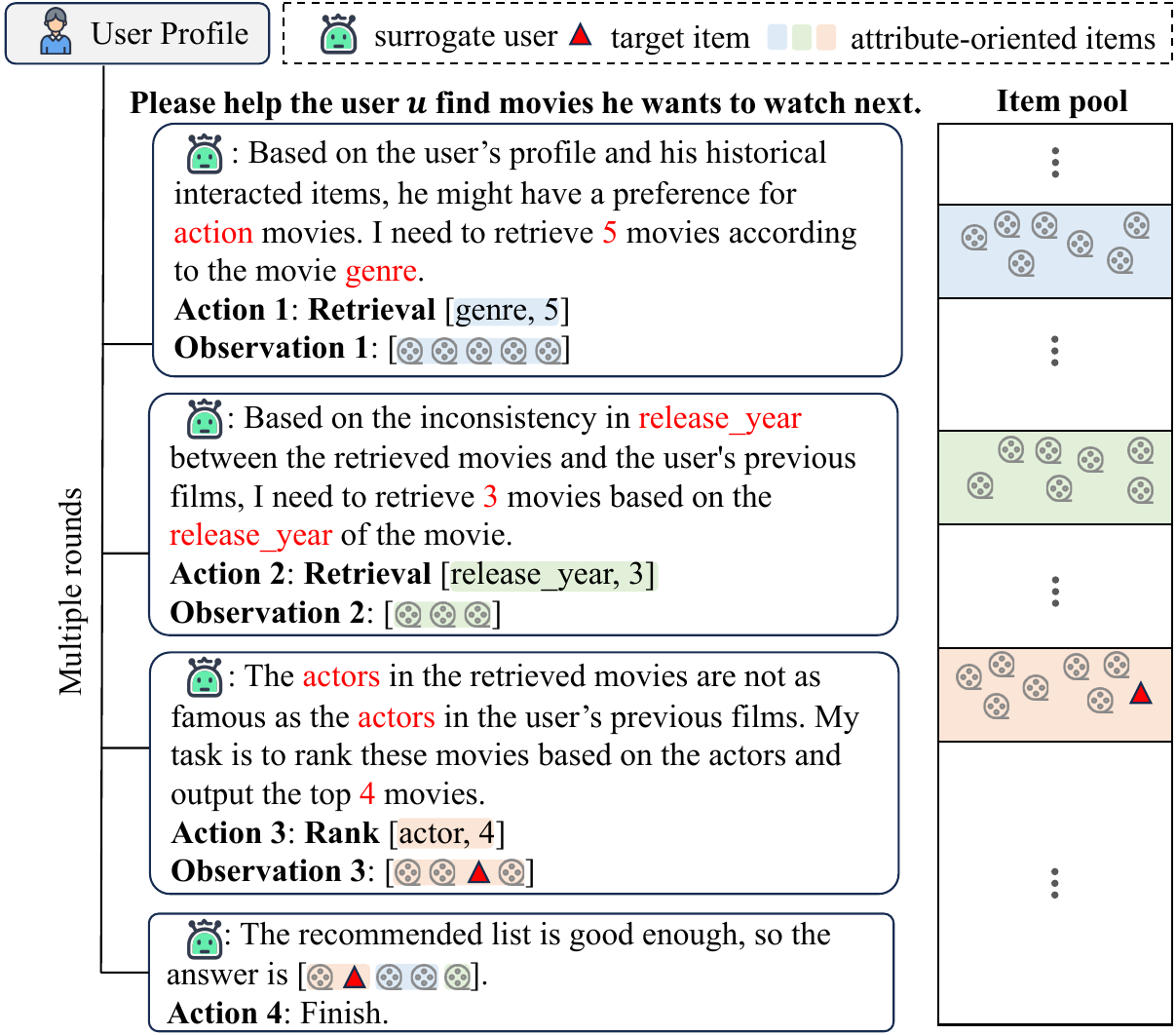}
\caption{Illustrating how \ours\ works. 
The LLM-based \textit{surrogate user} learns the real user's preferences and decides to employ attribute-oriented tools to explore  areas of items. This process leads to a broad view of items, which, in turn, leads to the successful retrieval of the target item. It is important to note that the areas representing different attribute-oriented items that are retrieved according to a specific attribute may contain overlapping elements.
}
\label{fig_intro}
\end{figure}

Inspired by the commonsense reasoning and knowledge utilization capabilities of large language models (LLMs), there have been several attempts to integrate LLMs with RSs and mitigate their inherent limitations~\cite{llmrec1, llmrec2, llmrec3}: 
\begin{itemize}[leftmargin=*,nosep]
\item \textbf{LLMs as RSs}: Here, LLMs, whether initially trained or further fine-tuned using user-item interaction data, are adapted to serve as RSs~\cite{zeroshot, mysore2023large, 23KAR} and directly generate candidate items in text. This approach easily suffers from the hallucination problem~\cite{llmHallucionation}, especially given large item catalogue sizes and extensive item names~\cite{InteRecAgent}. 
\item \textbf{LLMs enhance RSs:} Here, RSs are enhanced with world knowledge and reasoning abilities of LLMs~\cite{zeroshot, mysore2023large, 23KAR, 21unbert, 21ubert, llara}. This category limits LLMs to offering semantic information within a conventional recommendation paradigm, sometimes leading to inconsistencies between the semantic and behavior spaces.
\item \textbf{LLMs control RSs:} Here, LLMs are used to monitor and control the recommendation pipeline.
Existing controllers either have simple control strategies~\cite{chatrec, llm4rank} or necessitate active user involvement~\cite{controll2}. Their decisions are rarely human-like, which hinders their effectiveness in applications.
\end{itemize} 

\header{Tool learning for recommendations}
Motivated by recent advancements in tool learning with foundation models~\citep[e.g.,][]{toolformer, react, hugginggpt}, we propose to use LLMs as a surrogate user to emulate her/his decision-making process along with the utilization of tools. 
At the core of our proposal is the task of learning to adaptively select appropriate recommender tools and curate a user-centric item list that is aligned with the user's preferences. 
Figure~\ref{fig_intro} illustrates a four-round example of how \ours\ works. The LLM starts the simulation by focusing on the movie genre and selects an initial set of 5 movies; satisfied with the genre of the returned movies, it then aims to complement the set based on the release\_year, and retrieves 3 additional movies; subsequently, the LLM refines its focus towards the actors, leading to an adjustment of four movies in the list.
Such iterative refinements continue until the simulator deems the movie list satisfactory enough to include the item of interest, thus finishing the recommendation process.
Notice how the LLM directs the recommendation through multiple decision-making rounds with attribute signals, contrasting with conventional RSs.
By adopting this approach, we aim to move beyond relying solely on users' historical interactions, resulting in a more tailored set of recommended items~(\cf the red area on the right of Figure \ref{fig_intro}).

Using LLM tool learning to simulate users' decision-making processes presents distinct challenges.
The first challenge concerns the recommendation ability of LLMs. 
Although LLMs are pretrained on extensive datasets~\cite{instructGPT, anthropic}, improving their ability to produce quality recommendations remains a challenge, particularly in domain-specific scenarios.
The second challenge is developing appropriate attribute-oriented tools. 
Attribute-oriented tools should not only be capable of exploring facets of the item pool (\eg~ genre, release year of movies) but also need to be effective in handling different attribute choices during decision simulation.
Lastly, leveraging LLMs to refine the set of candidate items in each round presents a significant challenge. 
This step could ensure that the final results benefit from an LLM's open-world knowledge and are not limited by a single ``narrow expert.''

\header{A new proposal for tool learning for recommendations}
To address the challenges listed above, we introduce \ours, for LLM-empowered recommendations via tool learning, which is aimed at aligning the emergent abilities of LLMs with the demands of recommender systems. 
\ours comprises three key components:
\begin{enumerate*}[label=(\roman*)]
\item A \textbf{user decision simulation module}: We use LLMs initialized with user behavior history, acting as a \textit{surrogate user} to evaluate user preferences against the current scenario. 
\item \textbf{Attribute-oriented tools}: We develop two distinct sets of attribute-based tools: rank tools and retrieval tools. The ranking tools are operationalized by LLMs with attribute-oriented ranking instructions, while the retrieval tools are operationalized by merging a frozen backbone with additional fine-tuned attribute encoders. These tools are activated when the \textit{surrogate user} identifies unsatisfactory attributes, and then fetches corresponding candidate items. 
\item A \textbf{memory strategy}: This component checks the presence of intermediate results and stores them with associated tool marks, aiding the LLM in leveraging open-world knowledge to refine the final recommendation list.
\end{enumerate*}
The latter two components are governed by the \textit{surrogate user}'s decisions, and conclude the recommendation process once a satisfactory candidate item list has been found. 
\ours's iterative framework integrates LLMs into recommender systems while enhancing the quality of recommendations.

\header{Contributions}
Our main contributions are as follows:

\begin{itemize}[leftmargin=*,nosep]

\item We propose \ours, a framework that deploys LLMs to enhance recommendations via tool learning. It employs LLMs to closely emulate user preferences, thereby improving the accuracy of recommendations generated during user decision simulation.

\item To better meet the surrogate user's needs, we incorporate at\-tribute-oriented tools and a memory strategy. Those components address the challenge of effective item retrieval based on identified attributes, ensuring the recommendations are well-aligned with user preferences.

\item Experimental results on three real-world datasets demonstrate the effectiveness of \ours, especially in domains enriched by world knowledge.
\end{itemize}

\negskip
\section{Related Work}
We review tool learning with LLMs and the use of LLMs for recommendation.

\subsection{LLMs with Tool Learning }
There is a growing trend to employ LLMs to construct autonomous agents to achieve decision-making capabilities~\cite{hugginggpt, toolformer, agentEX1, toolllm, 23ghost}. 
These LLM-based agents often fall short in domains that demand extensive expert knowledge and suffer from hallucination issues~\cite{renminAgentsurvey}. 
To alleviate these problems, these agents are enhanced with the ability to invoke external tools for action execution. Previous external tools can be grouped into three types: APIs~\cite{hugginggpt, li2023api, toolllm, toolformer, toolSurvey}, Databases \& Knowledge Bases~\cite{22mrkl, chatdb, 23openagi}, and External Models~\cite{23memorybank, 23vipergpt, 23mmreact}.

The use of APIs as tools has become a popular approach. 
E.g., HuggingGPT~\cite{hugginggpt} employs models on HuggingFace to accomplish complex user tasks. 
API-Bank~\cite{li2023api} serves as an LLM-based API recommendation agent, and auto\-nomously searches and generates suitable API calls across various programming languages and domains. ToolBench~\cite{toolSurvey} is an LLM-based tool generation system that creates various tools based on natural language requirements.
Connecting LLMs to external databases or knowledge bases enables access to domain-specific information, thus generating more realistic actions. E.g., ChatDB~\cite{chatdb} uses SQL statements to query databases, enabling logical action execution by agents. 
Similarly, in the recommendation scenario, RecMind~\cite{23recmind} and InteRecAgent~\cite{InteRecAgent} engage various expert systems such as MySQL and planners to retrieve detailed item information.
Employing external models can expand the range of feasible actions.
E.g., MemoryBank~\cite{23memorybank} employs two language models to enhance text retrieval capabilities: one for encoding input text and the other for matching query statements. MM-REACT~\cite{23mmreact} integrates various vision models to improve its performance in visual understanding tasks.

Our attribute-oriented rank tools can be summarized into APIs, while retrieval tools can be summarized into external models. This approach helps our surrogate user to explore different interest areas and return a better user preference-aligned item set.

\negskip
\subsection{LLMs for Recommendation }

Methods using LLMs in RSs come in three groups, based on the role of the LLMs: LLMs as RSs~\cite{p5, 23up5, 23vip5, tallrec, instructrec, xu2023openp5, Yang2023RecInterpret}, LLMs as assistants~\cite{zeroshot, mysore2023large, 23KAR}, and LLMs as pipeline controllers~\cite{chatrec, controll2, 23recmind, InteRecAgent}.

LLMs as RSs involve using LLMs to generate candidate items. For instance, P5~\cite{p5} is fine-tuned on T5~\cite{T5} with a collection of personalized prompts to achieve zero-shot generalization. Following this, UP5~\cite{23up5} and VIP5~\cite{23vip5} extend the paradigm to fairness and multimodal tasks, respectively. InstructRec~\cite{instructrec} views recommendation as an instruction following task for LLMs, tuning T5 with user-specific instructions. TALLRec~\cite{tallrec} trains LLaMA~\cite{23llama} with LoRA~\cite{21lora} to follow instructions and respond to a binary query provided within the contextual information. LLaRA~\cite{llara} utilizes hybrid prompting to bridge the modality gap between traditional recommendation systems and LLMs.

When LLMs serve as Assistants, LLMs are leveraged for their factual knowledge or reasoning capabilities to generate or encode auxiliary textual features, assisting conventional RSs with semantic information. For example, Various studies use auxiliary textual features from BERT~\cite{18bert} for document ranking~\cite{21preBaidu}, while others aim to leverage these features for news recommendation~\cite{21unbert, 21empoweringnews, 21ubert}. KAR~\cite{23KAR} extracts reasoning and factual knowledge from LLMs to serve as augmented features, enhancing recommendation models in a model-agnostic manner. 

In the setting where LLMs act as Pipeline Controllers, models such as Chat-REC~\cite{chatrec} use ChatGPT to understand user preferences, decide whether to use the backend recommender system, and refine the suggested items before showing them to the user. RecLLM~\cite{controll2} and InteRecAgent~\cite{InteRecAgent} propose frameworks for integrated conversational recommender systems with LLMs managing dialogues, understanding user preferences. InteRecAgent further determines the usage of various tools (i.e., information query tools, retrieval tools, and rank tools) to support the candidate items. RecMind~\cite{23recmind}, inspired by P5's various-task experiment setup, manages tool usage through a Self-Inspiring mechanism. Agent4Rec~\cite{agent4rec} simulates and infers user-personalized preferences and behavior patterns using a LLM-based movie recommendation simulator.

While these pipeline controller efforts address challenges similar to ours, there are several aspects that make our approach different from theirs. First, we focus on the sequential recommendation task and top-N recommendation setting. Second, we introduce attribute-oriented tools designed specifically for the recommendation exploration journey. Lastly, we propose a proactive analysis of preference mismatches by using an LLM as a surrogate user.

\negskip
\section{Methodology}
\label{sec:method}

\begin{figure}[t]
\includegraphics[width=0.45\textwidth]{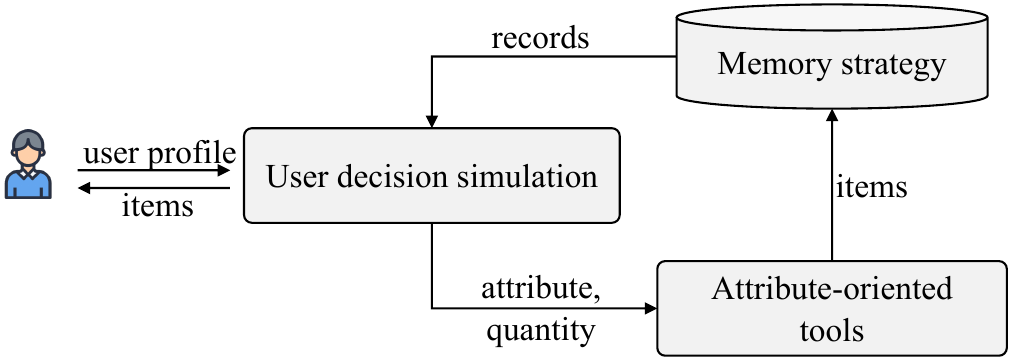}
\caption{An overview of the proposed LLM-based recommendation method via tool learning.
}
\label{meth_fig1}
\end{figure}

We propose \ours, as illustrated in Figure~\ref{meth_fig1}. 
We first formalize the recommendation process as a process exploring users' interests. 
Then, we introduce a general framework that adapts LLMs as surrogate users to enhance the recommendation mechanism through attribute-oriented tools. 

\negskip
\subsection{Problem Formulation}
In the context of sequential recommendation, as illustrated in Figure \ref{fig_intro}, given user $u$ with a historical interaction sequence $\mathcal{H} = \{i^1, i^2, \ldots, i^{n-1}\}$, the goal is to predict the next item of interest $i^n \in \mathcal{I}$, where $\mathcal{I}$ is the complete item pool. Conventional RSs retrieve a subset $\mathcal{I}_{c} \subseteq \mathcal{I}$ based on the predicted preferences of user $u$. However, this easily results in the desired item $i^n$ being absent from $\mathcal{I}_{c}$.

In this work, we position LLMs as the central controller for recommendation, simulating the user exploration \wrt item attributes in a multi-round manner. The interaction history $\mathcal{H}$ of user $u$ is fed into the LLM, so as to initialize the profile of surrogate user $\hat{u}$. 
In the first round, $\hat{u}$ identifies a key attribute $a_1$ and uses external tools to fetch the related item set $\mathcal{I}^{a_1}_{\hat{u}}$. 
For clarity and brevity, we omit the subscript $\hat{u}$ when the user is clear from the context. In the second round, $\hat{u}$ contrasts the preferences between $\mathcal{I}^{a_1}$ and $\mathcal{H}$, selects another attribute $a_2$, and retrieves $\mathcal{I}^{a_2}$. 
Such iterative refinements continue until $\hat{u}$ is satisfied with the retrieved items, and outputs the final set $\mathcal{I}_{\hat{u}}$ ranked by preference. Each derived item set reflects the interest emerging from the respective round.

\subsection{User Decision Simulation}\label{meh:1}
To validate LLMs' capability in simulating user preferences and utilizing external tools, we draw inspiration from prior tool-learning studies~\cite{react, toolSurvey, toolformer, toolllm} and propose a user decision simulation process. Here is an example of the simulation prompt in the movie recommendation scenario:

\begin{figure}[h]
\includegraphics[width=1\columnwidth]{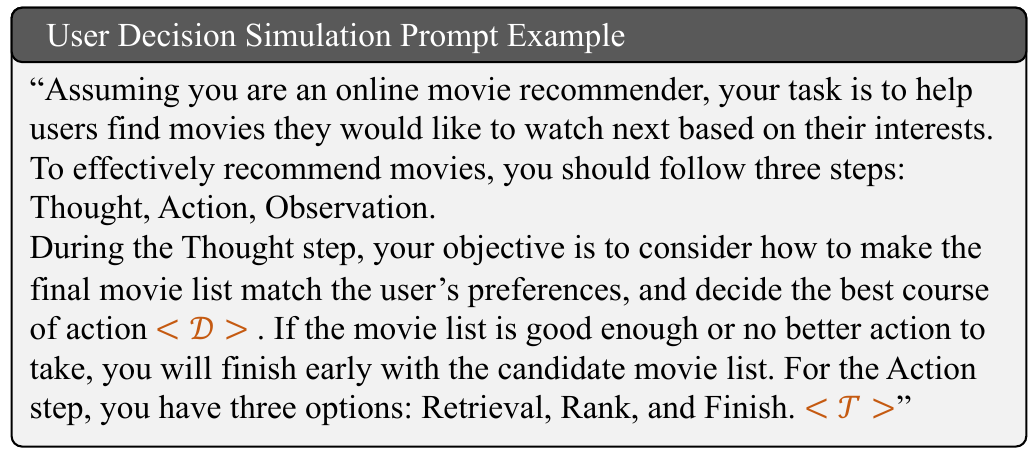}
\end{figure}

\smallskip\noindent%
where ${<}\mathcal{D}{>}$ represents the collection of demonstrations that show LLMs what a good movie list looks like, and ${<}\mathcal{T}{>}$ denotes the detailed description of the tools. 
The foundational elements of the user decision simulation methodology are twofold:
\begin{description}[leftmargin=*] 
    \item[Chain-of-thought prompting (CoT).] \label{meh:coT}  We use CoT~\cite{chainofthought} to synergize reasoning and action. 
    Here, ``reasoning'' refers to the ``thinking procedure'' for how to recommend suitable items to users, and decide the details of the subsequent action. 
    Meanwhile, ``action'' entails executing the directives derived from reasoning, making ``observations,'' and yielding the corresponding outcome. 
    At each step $t$, the LLM-driven surrogate user $\hat{u}$ receives an observation $o^{t-1}$ from the last step and derives a thought $g_{t}$ to take an action $\mathcal{A}_{t}$ following some policy $\pi(g_{t}, \mathcal{A}_{t}|c_{t})$, where $\pi$ could be implemented by any LLMs, specifically ChatGPT, in our \ours, and $c_{t} = (o_0, g_1, \mathcal{A}_1, o_1, \ldots, g_{{t}-1}, \mathcal{A}_{{t}-1}, o_{{t}-1})$ is the \textit{context} to the $\hat{u}$.
    The primary objective is to deduce the policy and mapping $c_{t} \rightarrow (g_{t}, \mathcal{A}_{t})$. 
    As shown in the left of Figure \ref{fig_intro}, within the context $c_{t}$, $\hat{u}$ observes a discrepancy in the movie actors between the retrieved movies and the user's previous movies. 
    Consequently, it decides to rank the movie list based on their actors (\ie~ $\mathcal{A}_{t}$) and obtain $o_{t}$. If $\hat{u}$ is satisfied within the context (\ie~ $c_N$), by comparing the candidate items to the user's history, then $\hat{u}$ will conclude the process and present the final set of candidate items, denoted as $\mathcal{I}_{\hat{u}}$.

    \item[Tool learning.] \label{meh:toolLearning}  Tool learning technology refers to combine the strength of LLMs and specialized tools, as discussed in previous studies~\cite{toolSurvey}. 
    Here, our tools are activated by the generated action $\mathcal{A}_{t}$~\cite{toolformer, toolllm, react}. 
    For each context $c_{t}$,  the initial observation $o_0 = (\mathcal{H}, \mathcal{T}, \mathcal{D})$ is composed of the user $u$'s historical interactions $\mathcal{H}$, tool description $\mathcal{T}$, and demonstration $\mathcal{D}$. 
    The tools can be categorized into two types based on \(\mathcal{T}_{type}\), which can be either ``retrieve tools'' or ``rank tools.'' 
    The description $\mathcal{T}$ elucidates the impact of using the tools, and $\mathcal{D}$ offers practical demonstrations of their application. 
    Together, these components promote the effective utilization of external tools.
    Tool actions are formulated as $\mathcal{T}_{type}[a_{t}, \$K]$, where $\mathcal{T}_{type} \in \{Retrieval, Rank\}$, $a_{t}$ indicates the chosen attribute based on $\hat{u}$'s decision at time $t$, and $\$K$ specifies the number of items to be returned. 
    For instance, as shown in Figure \ref{fig_intro}, at the first step, $\hat{u}$ opts to use the retrieval tools conditioned on the attribute ``genre'' and retrieves 5 items.  Meanwhile, at step 3, $\hat{u}$ decides to employ rank tools conditioned on the attribute ``actor'', returning the top 4 items.
    This phase is constructed to emulate a human-like approach of leveraging tools to broaden their choices through linguistic reasoning.
\end{description} 
The CoT prompting phase controls the iterative decision process, determining when to employ external tools or finalize the recommendations. Concurrently, feedback from these tools enhances item exploration and further refines the recommendation process.

\subsection{Attribute-oriented Tools}\label{sec:meh:tools:R}
By empowering an LLM to use tools, we can explore different parts of the item pool, uncovering target items that remain latent. To achieve this, we have designed two types of tools.
\subsubsection{Rank tools}
For attribute-oriented rank tools, we incorporate a ranking instruction template and employ LLMs to order the candidate items. 
Given that LLMs have demonstrated proficiency in both zero-shot and few-shot scenarios~\cite{llm4rank}, their capabilities are essential for returning item sets that align more closely with the user's latent intent.
Our instruction for ranking is framed as: ``\{\textit{User Historical Record}\} \{\textit{Prior Retrieved Item Set}\}. \textit{Please rank the above recommended movies by measuring the possibilities that the user would like to watch next most according to the movie} [\textit{Attribute Pattern} $a_{t}$] \textit{attribute and the given movie history records, and output top} [\textit{Output Size Pattern} $\$K$] \textit{movies except user's historical movies}.''

\subsubsection{Retrieval tools}
For each user, our attribute-oriented retrieval tools accept an attribute pattern $a_{t}$ and a specified item set size $\$K$, subsequently returning the matching candidate item set. An intuitive way is creating dedicated models tailored to each attribute pattern. However, it is markedly inefficient to fully train and store separate models for every possible attribute permutation. To address this, we introduce a two-stage method for managing attribute-specific variations:

\begin{description}[leftmargin=*]

\item[Pre-training.] The pre-training stage is for the original model with no attribute-specific consideration. Without loss of generality, we incorporate the attribute-specific switches into the popular sequential recommendation model, SASRec~\cite{sasrec}. 
For a historical behavior sequence $\mathcal{H}$, the $l$-th layer behavior representation matrix is denoted as $\mathbf{H}^l = \{\mathbf{h}_{1}^l, \mathbf{h}_{2}^l, \ldots, \mathbf{h}_{|\mathcal{H}|}^l\}$, where $\mathbf{h}_{i}^l$ is the $l$-th layer's representation of the $i$-th behavior within the sequence $\mathcal{H}$. Then in the final layer $L$, the pre-training behavior representation $\mathbf{H}$ is derived:
\begin{equation}
    \mathbf{H} = f_{seq}(\mathcal{H}|\beta) = \mathbf{h}_{|\mathcal{H}|}^L, \, \mathbf{H}^{l + 1} = \text{Transformer}^l_h(\mathbf{H}^{l}),
\end{equation}
where $f_{seq}$ represents the sequential model, $\beta$ is the pre-training parameter, $\text{Transformer}$ denotes the Transformer architecture encoder, and $\mathbf{H}$ is considered as the user representation at the pre-training phase for sequential recommendation.

\begin{figure}[t]
    \includegraphics[width=1\columnwidth]{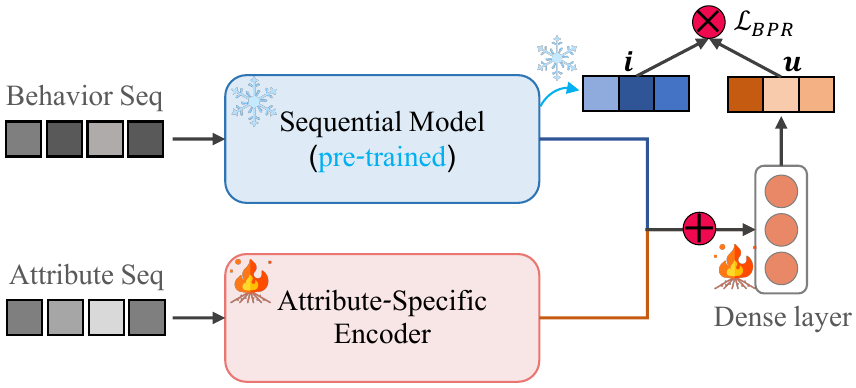}
    \caption{Fine-tuning stage of attribute-oriented retrieval tools. The parameters in the blue `ice' section remain fixed, while those in the red `flame' section are exclusively fine-tuned.}
    \label{meth_fig2}
\end{figure}

\item[Attribute-specific encoder in tuning.] As illustrated in Figure~\ref{meth_fig2}, after learning the pretrained model, we freeze the pre-training parameter $\beta$. Our goal is to fine-tune the attribute-specific encoder. To achieve this, we construct an additional attribute encoder $f_{attr}$. This encoder takes the user's historical item attribute sequence, $a_{u}$, as input. Importantly, $\mathcal{H}$ and $a_u$ are aligned based on items in the historical sequence. Subsequently, we define the $l$-th layer attribute representation matrix as  $\bm{a}_{u}^l = \{\bm{a}_{1}^l, \bm{a}_{2}^l, \cdots, \bm{a}_{|a_{u}|}^l\}$. Similar to the pre-training approach, the attribute sequence representation $\bm{a}_{u}$ is learned as:
\begin{equation}
    \bm{a}_{u} = f_{attr}(a_{u}|\gamma) = \bm{a}_{|a_u|}^L, \, \bm{a}_u^{l + 1} = \text{Transformer}^l_a(\bm{a}_u^{l}).
\end{equation}
We then incorporate a dense layer to encode the combination of attribute representation $\bm{a}_{u}$ and the frozen behavior representation $\mathbf{H}$, resulting in a new user representation:
\begin{equation}
    \mathbf{u} = \text{Dense}(\bm{a}_{u} \oplus \mathbf{H}, \theta),
\end{equation}
where $\gamma$ and $\theta$ are the trainable parameters in the tuning phase.
\end{description} 

\smallskip\noindent%
The sequential recommender is trained by minimizing the BPR loss function:
\begin{equation}
    \mathcal{L}_{BPR} = - \sum_{(\mathcal{H},\, v) \in \mathcal{O}^+, (\mathcal{H}, w) \in \mathcal{O}^-,} \log \sigma(\phi(\mathbf{u}, v) - \phi(\mathbf{u}, w)),
\end{equation}
where $\mathcal{O}^+$ and $\mathcal{O}^-$ denote the positive samples and negative samples, $\phi(\cdot)$ represents the inner-product layer, and $\sigma(\cdot)$ refers to the Sigmoid activation function. During the pre-training phase, $\mathbf{u}$ is represented by $\mathbf{H}$. The finetuned user embedding $\mathbf{u}$ is designed to be sensitive to the specific attribute, while maintaining the personalized sequential recommendation learned in the pre-training phase.

\subsection{Memory Strategy} \label{sec:method4}
The vast number of items and the complex item names\&IDs pose a challenge for LLMs when generating control commands or tool usages. Additionally, items retrieved from various tools should be systematically ordered to aid the LLM-based surrogate user $\hat{u}$ in making decisions. Therefore, we introduce a memory strategy, ensuring the correctness of generated items and cataloging candidate items with their respective tool annotations. 

The memory strategy is initialized with the item pool directory. Whenever external tools return candidate items, particularly from attribute-oriented rank tools, the strategy verifies the presence of these items in the initial directory. If there are any discrepancies, the tools are prompted to re-run with additional incorrect details attached behind. Once validated, the candidate items are recorded alongside their associated tool marks, in order to serve the subsequent tool calls. As an illustration, a typical prompt might be ``\textit{Here's the top} [\textit{Output Size Pattern} $\$K$] \textit{movie ID, movie name, and the recommendation confidence score from the recommender system with} [\textit{Attribute Pattern} $a_{\hat{u}}$] \textit{type}. \{\textit{Candidate Item Set}\}.''

\section{Experiments}
\label{exper}

In this section, we report on extensive experiments aimed at evaluating the performance of our proposed \ours. 
Our experiments focus on answering the following research questions:
\begin{enumerate*}[label=\textbf{(RQ\arabic*)},leftmargin=*]
\item How does \ours~ compare to conventional RSs and LLM-based RSs in the sequential recommendation setting?

\item How do different components 
(\textit{\ie~ \textit{user decision simulation}, \textit{termination round}, \textit{attribute-oriented retrieval tools}}) 
influence our \ours? And

\item Are LLMs capable of using their inherent knowledge to cater to the recommendation task's needs?
\end{enumerate*}

\subsection{Experimental Settings}
\label{exper:setting}

\subsubsection{Datasets}
To evaluate the effectiveness of our methods, we conduct experiments on three real-world datasets: Ml-1M, Amazon-Book, and Yelp2018.
\begin{itemize}[leftmargin=*]
\item \textbf{ML-1M}. This dataset is derived from the MovieLens-1M\footnote{\url{https://grouplens.org/datasets/movielens/}} benchmark, which contains user ratings for movies with timestamps. 
We take the movies' \textit{genre} and \textit{release year} as the attribute information.

\item \textbf{Amazon-Book}. This dataset\footnote{\url{https://nijianmo.github.io/amazon/}} is extracted from \textit{Amazon.com} platform. We adopt the Book category to evaluate our method.  We take the books' \textit{Price} and \textit{Sales rank} as the attribute information. A 10-core setting is applied to maintain dataset quality.

\item \textbf{Yelp2018}. This dataset is collected from the 2018 edition of the Yelp  Challenge.\footnote{\url{https://www.yelp.com/dataset}} We utilize local businesses' \textit{Categories}, \textit{City} and \textit{Stars} as attributes. We employ the 10-core setting to ensure a minimum of ten interactions for each user and item.
\end{itemize}
For each dataset, we organize users' interactions chronologically based on timestamps, allowing us to create the corresponding historical interaction sequences. Items are described using their product IDs\&Names. We summarize the statistics of our datasets in Table \ref{tab:dataset}.

\begin{table}[!t]
\caption{The statistics of the datasets used.}
\begin{tabular}{l rrrc}
\toprule
Datasets     & \multicolumn{1}{c}{\#Users} & \multicolumn{1}{c}{\#Items} & \multicolumn{1}{c}{\#Interactions} & \multicolumn{1}{c}{Sparsity (\%)}  \\ 
\midrule
ML-1M         & 6,041                      & 3,884                     & 1,000,209  & 95.74                                          \\
Amazon-Book & 158,349                       & 97,566                      & 3,876,663    & 99.97                                   \\
Yelp2018     & 77,278                      & 45,582                     & 2,102,836    & 99.94                              \\ 
\bottomrule
\end{tabular}
\label{tab:dataset}
\end{table}

\subsubsection{Evaluation protocols}
We apply the leave-one-out strategy~\cite{sasrec, 19bert4rec} and employ timestamps to set the sequence order, dividing the interaction data into training, validation, and test sets. The attribute-oriented retrieval tools are trained on the training and validation sets. 
To measure the recommendation performance, we adopt two widely used metrics \textit{NDCG@N} and \textit{Recall@N} to evaluate the results within the top-\textit{N} positions, with $N=10$ in our experiments.
Due to budget constraints, following prior work~\cite{chatrec, llm4rank}, we randomly sample 200 users and their historical behaviors from the test set for each dataset. 
A similar, and similarly-sized, setting has been adopted in other recent LLM-related recommendation researches~\cite{agent4rec, agentcf23}. 
To enhance the robustness and credibility of our results, we repeated the experiments three times, each with a different sample of 200 users. 
The average results and standard deviations from these trials are presented in Table~\ref{tab:main}.

\begin{table*}[!t]
    \caption{The test performance comparison on three real-world datasets. The bold font denotes the winner in that column.  The row ``Improvement'' indicates the relative performance gain of our \ours\ and the suboptimal method.}
    \label{tab:main}
    \begin{tabular}{l@{\hspace{1pt}}c@{\hspace{3pt}}cc@{\hspace{3pt}}cc@{\hspace{3pt}}c}
    \toprule
                  & \multicolumn{2}{c}{ML-1M} & \multicolumn{2}{c}{Amazon-Book} & \multicolumn{2}{c}{Yelp2018} \\  
                  \cmidrule(r){2-3}
                  \cmidrule(r){4-5}
                  \cmidrule{6-7}
                  & Recall      & NDCG          & Recall      & NDCG            & Recall      & NDCG      \\ 
                  \midrule
    SASRec        & 0.203$\pm$\text{\scriptsize 0.047}      & \underline{0.1017}$\pm$\text{\scriptsize 0.016}      & 0.047$\pm$\text{\scriptsize 0.015}            & 0.0205$\pm$\text{\scriptsize 0.006}         & 0.030$\pm$\text{\scriptsize 0.005}         & 0.0165$\pm$\text{\scriptsize 0.006}       \\
    BERT4Rec      & 0.158$\pm$\text{\scriptsize 0.024}        & 0.0729$\pm$\text{\scriptsize 0.008}      & 0.042$\pm$\text{\scriptsize 0.015}           & 0.0212$\pm$\text{\scriptsize 0.009}        & \underline{0.033}$\pm$\text{\scriptsize 0.021}          & \textbf{0.0218}$\pm$\text{\scriptsize 0.016}        \\
    P5       & \underline{0.208}$\pm$\text{\scriptsize 0.021}       & 0.0962$\pm$\text{\scriptsize 0.009}      & 0.006$\pm$\text{\scriptsize 0.003}           & 0.0026$\pm$\text{\scriptsize 0.002}        & 0.012$\pm$\text{\scriptsize 0.005}         & 0.005$\pm$\text{\scriptsize 0.001}       \\ 
    SASRec$_{BERT}$   & 0.192$\pm$\text{\scriptsize 0.015}       & 0.0967$\pm$\text{\scriptsize 0.006}      & 0.042$\pm$\text{\scriptsize 0.003}           & 0.0194$\pm$\text{\scriptsize 0.002}        & 0.032$\pm$\text{\scriptsize 0.016}         & 0.0131$\pm$\text{\scriptsize 0.007}       \\
    BERT4Rec$_{BERT}$ & 0.202$\pm$\text{\scriptsize 0.013}       & 0.0961$\pm$\text{\scriptsize 0.009}       & 0.045$\pm$\text{\scriptsize 0.023}           & 0.0233$\pm$\text{\scriptsize 0.012}         & \textbf{0.040}$\pm$\text{\scriptsize 0.028}          & \underline{0.0208}$\pm$\text{\scriptsize 0.015}       \\
    Chat-REC      & 0.185$\pm$\text{\scriptsize 0.044}       & 0.1012$\pm$\text{\scriptsize 0.016}      & 0.033$\pm$\text{\scriptsize 0.015}            & 0.0171$\pm$\text{\scriptsize 0.007}        & 0.022$\pm$\text{\scriptsize 0.003}          & 0.0121$\pm$\text{\scriptsize 0.001}       \\
    LLMRank       & 0.183$\pm$\text{\scriptsize 0.049}       & 0.0991$\pm$\text{\scriptsize 0.020}      & \underline{0.047}$\pm$\text{\scriptsize 0.013}           & \underline{0.0246}$\pm$\text{\scriptsize 0.004}        & 0.030$\pm$\text{\scriptsize 0.005}         & 0.0140$\pm$\text{\scriptsize 0.004}       \\ 
    \midrule
    \ours      & \textbf{0.215}$\pm$\text{\scriptsize 0.044}       & \textbf{0.1171}$\pm$\text{\scriptsize 0.018}      & \textbf{0.053}$\pm$\text{\scriptsize 0.013}            & \textbf{0.0259}$\pm$\text{\scriptsize 0.005}        & 0.028$\pm$\text{\scriptsize 0.003}          & 0.0159$\pm$\text{\scriptsize 0.001}       \\
    ToolRec$_{B}$      & 0.185$\pm$\text{\scriptsize 0.018}        & 0.0895$\pm$\text{\scriptsize 0.002}      & 0.043$\pm$\text{\scriptsize 0.013}            & 0.0223$\pm$\text{\scriptsize 0.008}        & 0.025$\pm$\text{\scriptsize 0.005}         & 0.0136$\pm$\text{\scriptsize 0.009}       \\ 
    \midrule
    Improvement         & 3.36\%      & 15.10\%     & 14.28\%         & 5.14\%        & --29.16\%      & --27.32\%      \\ 
    \bottomrule
    \end{tabular}
    \end{table*}

\subsubsection{Baselines}
\label{baseline_222}
We compare \ours\footnote{Our codes are available at~\url{https://github.com/Go0day/ToolRec-Code}.} against two traditional approaches (the first two below), one that uses LLMs as RSs (the following one), two that enhance RSs with LLMs (the next two), and two that use LLMs to control RSs (the remaining two).
\begin{itemize}[leftmargin=*,nosep]
\item \textbf{SASRec}~\cite{sasrec}. A self-attention-based sequential recommender, which employs the encoder of the Transformer architecture to generate representations of users' behavior sequences.

\item \textbf{BERT4Rec}~\cite{19bert4rec}. A bidirectional self-attention-based sequential recommender. It uses the Transformer encoder to predict randomly masked items in a sequence by conditioning on both their left and right context, thereby capturing user historical behaviors.

\item \textbf{P5}~\cite{p5}. An encoder–decoder Transformer-based approach that unifies different recommendation related tasks into a single generative LLM. 
For our sequential recommendation downstream task, we adopt the personalized prompts from OpenP5~\cite{xu2023openp5} and apply the same random indexing method to items as used in our \ours. This model is fully fine-tuned using these personalized prompts on the pre-trained T5-small~\cite{T5}.

\item \textbf{SASRec$_{BERT}$}~\cite{18bert}. An attention-based method that modifies the single interaction sequence encoder by adding a feature encoder (structured as shown in Figure \ref{meth_fig2}). This model is fully fine-tuned using semantic representations pre-trained with BERT.

\item \textbf{BERT4Rec$_{BERT}$}. A variation of the BERT4Rec sequential recommender enhanced with BERT's pre-trained representations.

\item \textbf{Chat-REC}~\cite{chatrec}. The first work was on using an LLM as a controller. For our sequential recommendation task, we adopt the recommendation prompt based on the original paper, choose SASRec to supply the candidate items, and modify the output format for parsing.

\item \textbf{LLMRank}~\cite{llm4rank}. A LLM-based ranking model. In our full-ranking experimental setup, we retrieve thirty candidate items (according to the original paper) from SASRec and adjust the output format for parsing.

\item \textbf{\ours}. The method that we propose. We primarily evaluate two versions of \ours: \ours that is implemented using external attribute-oriented retrieval tools with frozen $SASRec$ parameters; and ToolRec$_{B}$ that is developed using BERT4Rec as the backbone for external attribute-oriented retrieval tools.

\end{itemize}

\subsubsection{Implementation details}
We use the gpt-3.5-turbo-16k  (ChatGPT for short) model as our primary LLM within \ours. This model is responsible for parsing user preferences and assisting in tool learning. We retain the default hyperparameters of ChatGPT without modifications. 
To enable \ours to emulate user decisions, we incorporate decision demonstrations into the prompt for in-context-learning.
For attribute-oriented tools, retrieval tools are enhanced with corresponding additional item attributes specific to each dataset. Attributes are represented using word embeddings from GloVe~\cite{14glove}. These tools are developed on two widely recognized backbones: SASRec and BERT4Rec. Additionally, our rank tools are constructed with instructions on ChatGPT. We limit our decision processes to a maximum of eight rounds.

\subsection{Performance Comparison (RQ1)}

We compare our proposal with conventional sequential recommenders and LLM-enhanced sequential recommenders, as detailed in Section \ref{baseline_222}. The results are reported in Table \ref{tab:main}, from which we observe:

\begin{itemize}[leftmargin=*,nosep]
\item  In general, \ours~ outperforms all baselines on the ML-1M and Amazon-Book datasets. 
The performance improvement can be attributed to the efficacy of our designed \ours~ framework. By delving into various facets of the item pool and utilizing LLMs to guide the exploration process, it can more effectively align with the user's intent.

\item \ours and ToolRec$_{B}$ consistently demonstrate improved performance compared to their respective underlying models (\ie~ SASRec and BERT4Rec).
This confirms the superiority of the \ours framework and underscores its adaptability.
The results suggest that our approach, which harnesses LLMs for recommendation through tool learning, has the potential to be integrated as a supplementary module in various recommendation systems.

\item \ours exhibits subpar performance on the Yelp2018 dataset. We attribute this to the LLM's limited knowledge of local businesses. Since LLMs are primarily trained on widely available web data, they might possess a more robust understanding of topics like movies and books than local (niche) businesses. 
In more specialized domains, \ours, which involves multiple interactions between the LLM and external tools, has the potential to make more incorrect decisions and may exhibit heightened deficiencies when compared to other LLM-based approaches such as LLMRank and SASRec$_{\text{BERT}}$.

\item In the approach where LLMs serve as RSs, P5 demonstrates strong performance on the ML-1M. This is exciting, considering we use datasets on the scale of millions~(Table \ref{tab:dataset}), and accurately generating items for users is challenging. However, P5 shows weaker performance on the Amazon-Book and Yelp2018 datasets, potentially due to the fact that the sparsity of those datasets is higher than ML-1M, thus leading to more severe hallucination issues. The development of more carefully designed item indexing methods~\cite{hua2023index} may help alleviate this problem, but this falls beyond the scope of our current research.

\item For LLM-enhanced RS approaches, such as SASRec$_{BERT}$ and BERT4Rec$_{BERT}$, it is not surprising to see improvements over their backbone models in most cases. However, on the ML-1M dataset,  SASRec$_{BERT}$ performs much worse than SASRec. Meanwhile, on the Amazon-Book dataset, BERT4Rec$_{BERT}$'s performance is comparable to its baseline version. While language models can generally augment recommendation tasks, addressing the disparity between semantic and behavioral spaces remains challenging.

\item For LLM-controlled RS approaches, such as Chat-REC and LLMRank, they couldn't achieve consistent improvements over SASRec in the same way as \ours. This suggests that merely using basic control strategies or employing LLM as a ranker might not be the most effective way forward for recommendation tasks. 
A more comprehensive strategy, as demonstrated by \ours, appears better suited to harnessing LLMs to enhance recommendations.

\end{itemize}

\subsection{Decomposing \ours~ (RQ2)}
Next, we delve deeper into \ours. We examine the user decision simulation, revealing how multi-round interactions enhance recommendation quality. We evaluate the efficiency of our attribute-oriented retrieval tools, highlighting the balance between rich information and computational practicality.


\subsubsection{Effectiveness of user decision simulation}\label{sec:exp:up}
To verify the contribution of the user decision simulation component (\cf Section~\ref{meh:coT}), we conducted an ablation study considering three variants of \ours: 
\begin{enumerate*}[label=(\roman*)]
\item Disabling the CoT and tool learning components of \ours, we forced the LLM to rank the candidate items from SASRec and output the result. This variant is denoted as ``w/ single''; 
\item Unlike ``w/ single'', we had the LLM rank candidate items using both SASRec and the attribute-oriented retrieval tools, termed ``w/ multi''; 
\item We disable the CoT component, and instead, the LLM was instructed to generate all the steps of tool-calling at once and then strictly follow the execution plan. This setup is termed ``w/ Plan''. 
\end{enumerate*}
To ensure fairness in our comparisons, both the ``w/ single'' and ``w/ multi'' variants use the same number of candidate items: thirty items in total.

\begin{figure}[!t] 
	\centering 
	\subfloat[ML-1M.]{
		\centering 
		\includegraphics[ width=1\columnwidth]{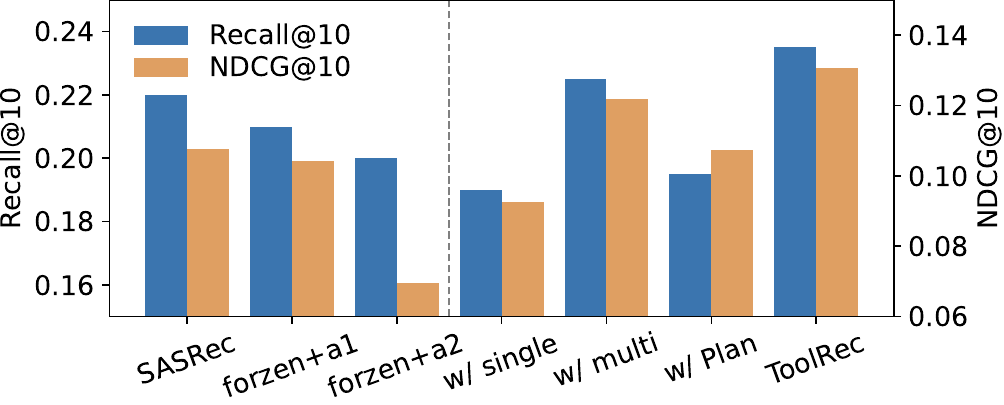}
	}
    \\[-.5mm]
	\subfloat[Amazon-Book.]{
		\centering
		\includegraphics[width=1\columnwidth]{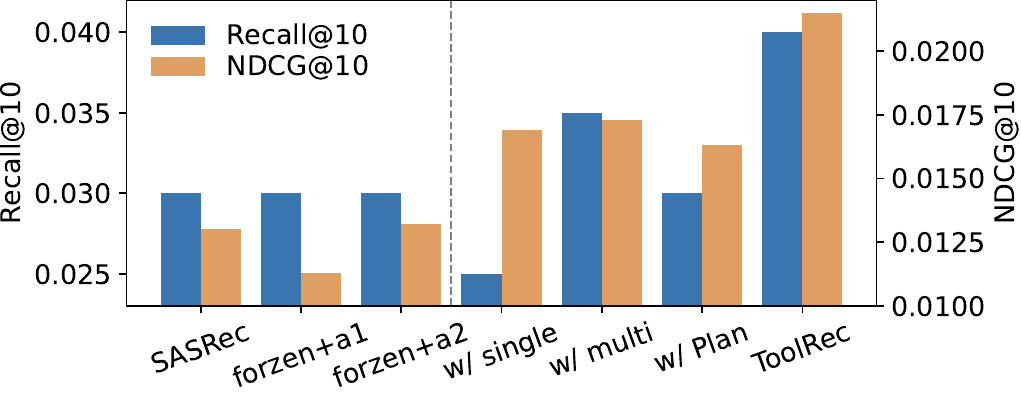}
	}
\caption{Performance of \ours\ and its variants. The right side of the dividing line indicating the methods involving LLMs.}
\label{fig:ablation}
\end{figure}

We demonstrate the experimental results in Figure \ref{fig:ablation} and have the following findings:
\begin{itemize}[leftmargin=*,nosep]
\item  Removing the CoT and tool learning component degrades the model's performance. The ``w/ single'' variant consistently underperforms both ``w/ multi'' and ``w/ Plan'', and its performance is even subpar compared to SASRec. 
This decline in performance can be attributed to ``w/ single'' relying exclusively on the LLM's zero-shot ranking without leveraging additional information to refine the results.

\item It is important to note that the performance of attribute-oriented retrieval tools, namely ``frozen + $a$1'' and ``frozen + $a$2'', is inferior to SASRec in dataset ML-1M. However, results refined by 'w/ multi' not only surpass 'w/ single' but also outperform SASRec. This improvement suggests that the strength of our approach is not solely due to the broader item size, but also derives from the assistance of the additional attribute information.

\item  When comparing ``w/ Plan'' with ``w/ multi'', the latter consistently achieves superior results. One potential reason might be that the ``generate-then-execute'' approach in ``w/ Plan'' lacks explicit performance guidance, causing it to miss the target. This further underscores the importance of the user decision simulation in our approach.
\end{itemize}

\begin{figure}[!t] 
	\centering 
	\subfloat[ML-1M.]{
		\centering 
		\includegraphics[ width=1\columnwidth]{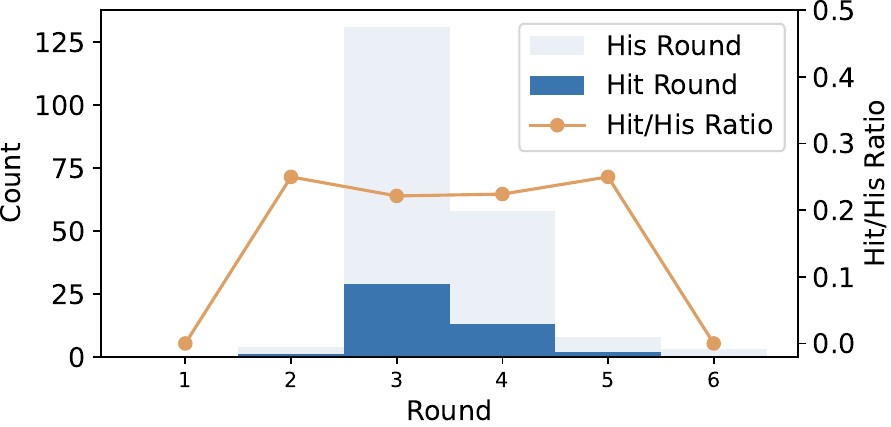}
	}
    \\[-.5mm]
	\subfloat[Amazon-Book.]{
		\centering
		\includegraphics[ width=1\columnwidth]{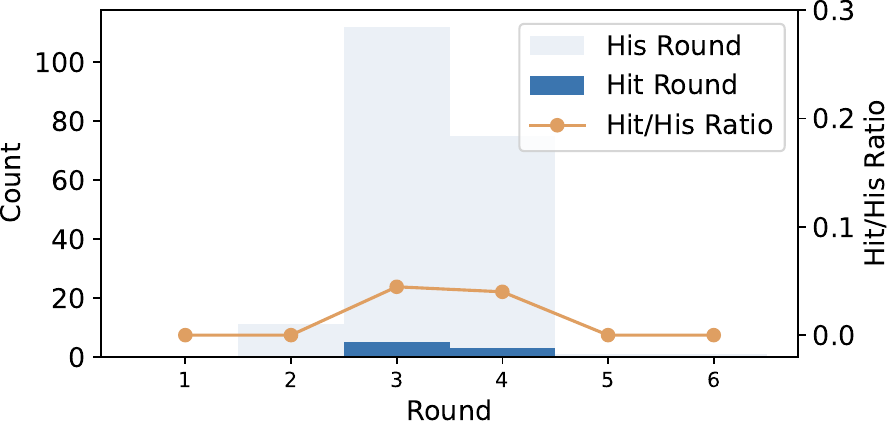}
	}
\caption{Distribution of termination rounds for \ours. ``His Round'' indicates the distribution of termination rounds for all users, while ``Hit Round'' highlights the termination round where the recommended list accurately contains the user's target item.}
\label{exp_round}
\end{figure}

\subsubsection{Analysis of round termination in \ours}
Figure \ref{exp_round} illustrates the distribution of the number of termination rounds, $N$, in \ours. 

Based on the data, we make the following observations:
\begin{enumerate*}[label=(\roman*)]
\item The majority of processes conclude within three or four rounds. This suggests that after a few iterations, our LLM-based surrogate user, denoted as $\hat{u}$, develops a good understanding of whether the user's preferences have been adequately addressed. 
\item While the majority of successful hits also occur within three or four rounds, an interesting trend emerges in the ML-1M dataset: both shorter and longer processes tend to be more successful in reaching the target item. One interpretation is that shorter rounds signify tasks that are more straightforward for the surrogate user $\hat{u}$. But for users with diverse interests or nuanced tastes, $\hat{u}$ might need additional rounds to gather more information and determine if the recommendation process has been satisfactorily completed.
\end{enumerate*}

\begin{figure}[!t] 
	\centering 
	\subfloat[ML-1M.]{
			\centering 
			\includegraphics[trim=0mm 0mm 0mm 2mm, width=1\columnwidth]{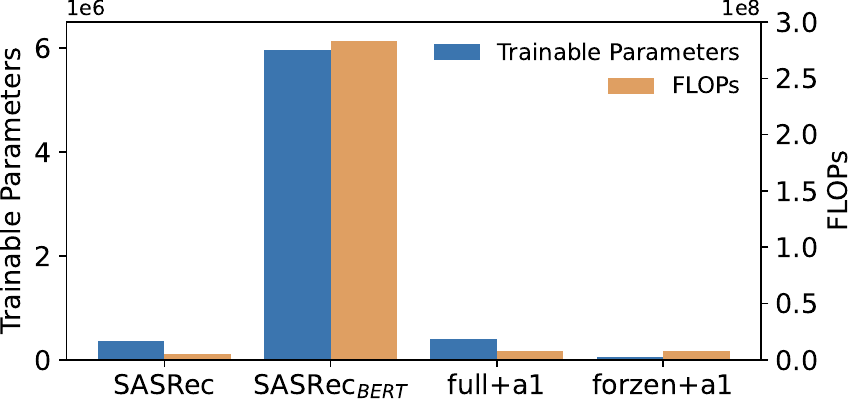}
	}
    \\
	\subfloat[Amazon-Book.]{
			\centering
			\includegraphics[trim=0mm 0mm 0mm 2mm, width=1\columnwidth]{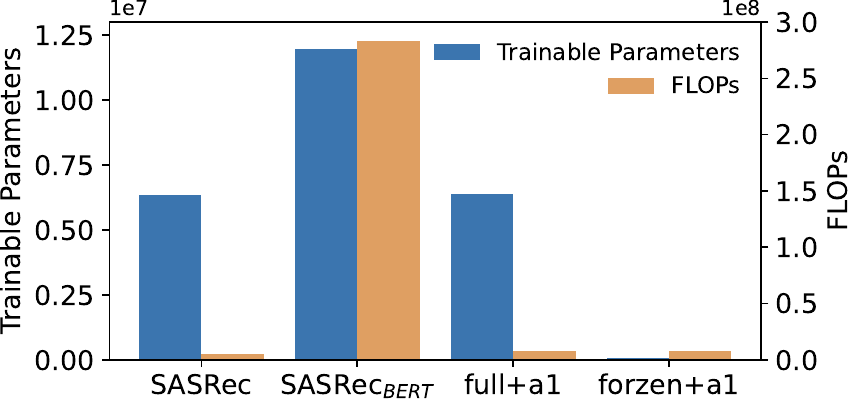}
	}
\caption{Comparison of trainable parameters and FLOPs for various retrieval model configurations.}
\label{exp_Retrieval}
\end{figure}

\vspace{-1mm}
\subsubsection{Efficiency and scalability of attribute-oriented retrieval tools}
As elaborated in Section \ref{sec:meh:tools:R}, the attribute-oriented retrieval tools are designed to adeptly follow diverse attribute choices.
Figure \ref{exp_Retrieval} compares the number of trainable parameters and FLOPs on the ML-1M and Amazon-Book datasets. Here, `forzen+$a$1' represents our attribute-oriented retrieval tool with frozen backbone parameters, while `full+$a$1' denotes its full fine-tuning variant.

We have the following observations:
\begin{enumerate*}[label=(\roman*)]
\item The SASRec$_{BERT}$ model is considerably larger than the original SASRec. This increase in size can be attributed to the enriched semantic information present in the BERT embedding.
\item The trainable parameter count for `full+$a$1' aligns closely with that of SASRec. Essentially, this is akin to training an entirely new model. Such an approach becomes impractical as the number of attributes escalates.
\item While both `forzen+$a$1' and `full+$a$1' exhibit identical FLOPs, the former boasts a significantly reduced count of trainable parameters compared to the latter. Moreover, since our `forzen+$a$1' employs the shared backbone of SASRec, storage is primarily reserved for the additive parameters from the attribute-specific encoder and the dense layer.
\end{enumerate*}



\subsection{Surprises and Limitations (RQ3)}
\ours benefits from LLMs to reason about users' preferences across attributes, enabling multi-round interaction processes and improved recommendations. 

However, language models are not inherently designed for recommendation tasks. 
Below, we discuss the failures and limitations this gives rise to.

\subsubsection{Unanticipated outcomes: Beyond conventional failures}
For a rigorous comparison with existing baselines, we employed the Memory strategy (\cf Section~\ref{sec:method4}). This strategy ensures that returned items align with the dataset directory, prompting a re-run if discrepancies arise. 

However, beyond the traditional recommendation paradigm, some samples that are typically classified as ``failures'' deserve a second look.

\begin{figure}[tb]
\includegraphics[width=1\columnwidth]{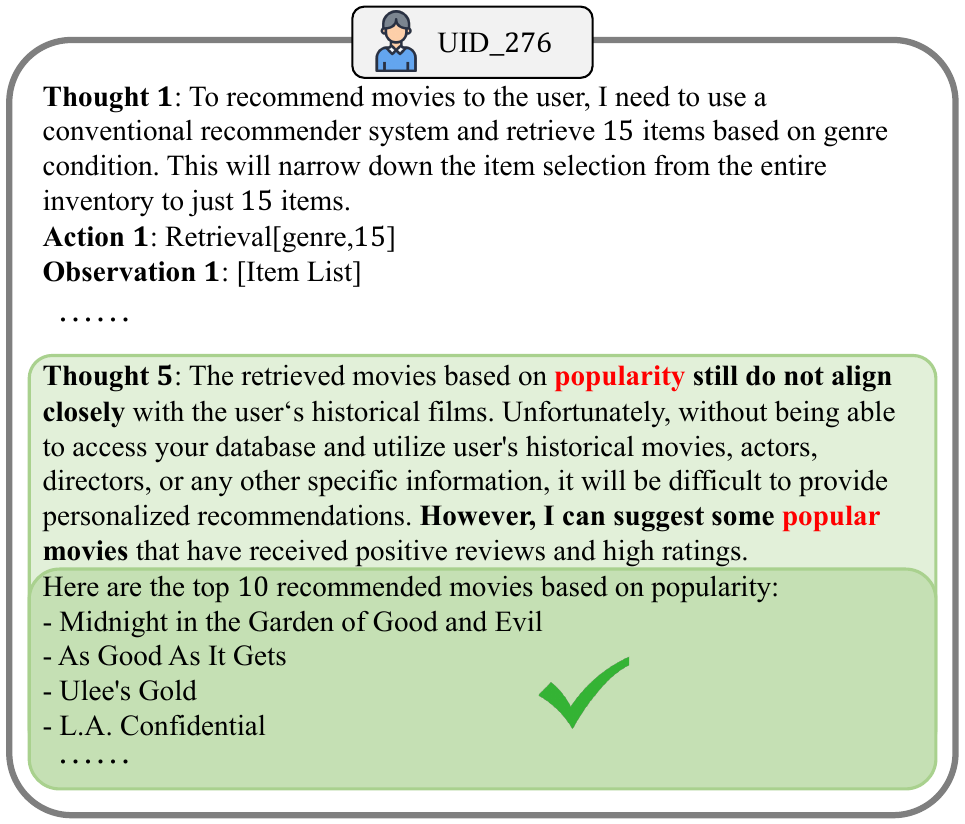}
\caption{Case study of parsing errors in output samples. Highlighted samples have been manually verified to exist in the real world.}
\label{exp_case}
\end{figure}

Figure~\ref{exp_case} shows an example. At \textit{Thought 5}, when our LLM-based surrogate user, $\hat{u}$, is unsatisfied with the retrieved movies, it typically evaluates the unsatisfactory attributes and uses external tools to uncover additional options, thereby refining the recommendation.

Yet, in this scenario, rather than using external tools or settling for the current list of candidates, $\hat{u}$ decides to suggest ``some popular movies'', providing a top-$N$ recommendation on its own. Since we never trained the LLM on any of our recommendation datasets, it is evident that not all of the movies it suggests are found in the ML-1M dataset. As a result, this action is labeled as a ``failure'', prompting a re-run. However, from another perspective, this could be viewed as a successful recommendation; the only limitation is our inability to evaluate it within the current dataset constraints.

\begin{figure}[!t] 
	\centering 
	\subfloat[ML-1M.]{
			\centering 
			\includegraphics[width=0.90\columnwidth]{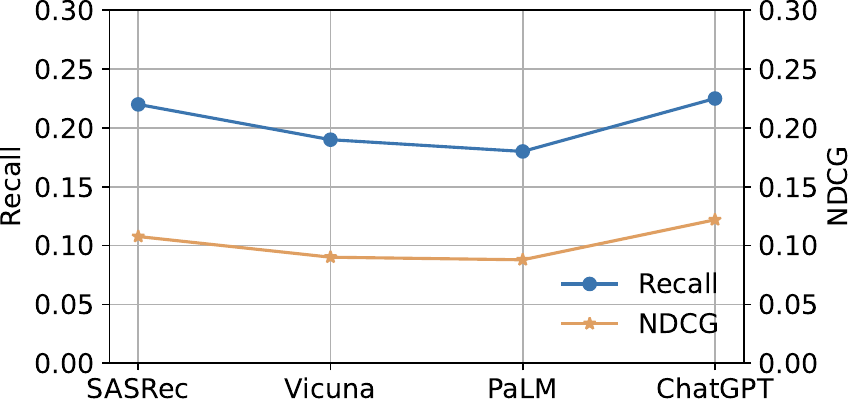}
	}
    \\[-.5mm]
	\subfloat[Yelp2018.]{
			\centering
			\includegraphics[width=0.90\columnwidth]{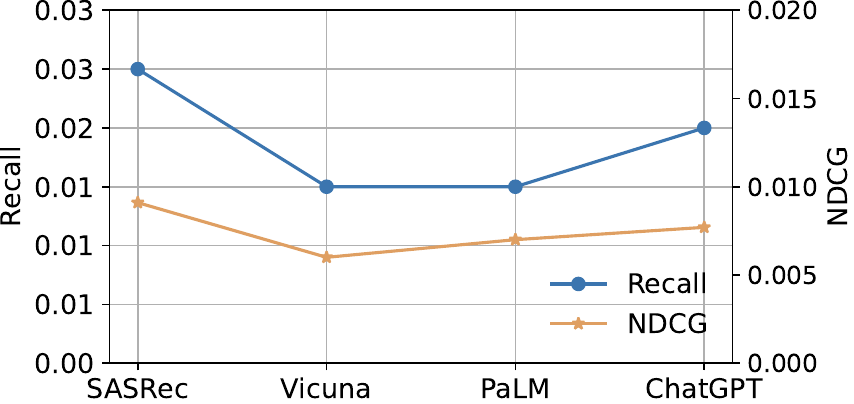}
	}
\caption{Ranking performance across different LLMs.}
\label{exp_rank}
\end{figure}

\subsubsection{Influence of LLM selection on recommendation performance}
Following the experimental setup described in Section~\ref{exper:setting}, we replace the base LLM with two alternative LLMs: Vicuna1.5-13B-16k~\cite{vicuna2023} (Vicuna in short), an open-source chatbot fine-tuned on Llama2~\cite{llama2}; and PaLM~\cite{22palm}, a commercial LLM developed by Google AI. 
However, both variants yielded subpar outcomes, either failing to adhere to task instructions, misconstruing tool usage, or misinterpreting user preferences (some representative failures are cataloged in the Appendix).
Notably, even though Vicuna and PaLM use the same prompt template as ChatGPT, they are unable to consistently generate recommendations across all test users, despite several attempts. 
This suggests that ChatGPT has better reasoning capabilities than both Vicuna and PaLM.

Beyond reasoning, a ranking experiment was conducted (\cf Section \ref{sec:exp:up}, w/ multi) to examine the open-world knowledge of various LLMs. 
As revealed in Figure \ref{exp_rank}, on datasets like ML-1M (and similar outcomes on Amazon-Book), both Vicuna and PaLM underperformed SASRec, whereas ChatGPT exhibited superior results. These findings align with insights presented in LLMRank~\cite{llm4rank}.
However, all three models lagged behind SASRec on the Yelp2018 dataset, suggesting a possible limitation of LLMs in contexts like local businesses, where open-world knowledge is limited and of more limited use than on the other datasets.

\section{Conclusion}
\label{sec:conclusion}

In this work, we zoomed in on the tool-learning capacities of LLMs, using them as controllers to guide the exploration of item spaces in a recommendation scenario. 
Specifically, by treating LLMs as surrogate users, they can adeptly capture the nuances of a current context alongside user preferences. 
Subsequently, we employ attribute-oriented tools for precise item retrieval. 
We developed two types of attribute-oriented tools: rank tools and retrieval tools, each fetching the corresponding candidate items. 
To enhance accurate item retrieval, items that appear in the process are verified and stored using the memory strategy.
Extensive experiments on real-world datasets rich in knowledge demonstrate the effectiveness and rationality of \ours.

The idea of using LLMs for simulation, either under the hood as in our approach or for counterfactual explorations while interacting with users, holds great potential for combining the strengths of LLMs and recommendation models. 
In the short term, companies operating their own recommender systems may find it impractical to switch to LLM-based RSs. 
However, \ours\ could enhance recommendation performance by integrating LLMs with their current systems. 
In the long-term, users are increasingly relying on LLMs for various daily tasks, including recommendations. 
\ours\ does not require extensive fine-tuning of the LLM, which can lead to additional costs and potential delays due to outdated information. 
Furthermore, the results from \ours\ are more reliable than those from zero-shot or few-shot LLM recommendations, as they are augmented by traditional recommendation system outputs.

Achieving strong recommendation performance hinges on a dataset with rich semantic knowledge and the robust capabilities of the LLM. 
In future work, we plan to incorporate recommendation knowledge into LLMs to enhance domain-specific tool learning, such as retrieval-augmented generation or fine-tuning LLMs, potentially reducing the reliance on rich semantic knowledge.  Additionally, we intend to explore different types of tools, including search engines and databases, along with a self-reflection strategy, to achieve even more personalized recommendations.

\subsubsection*{\bf Acknowledgments.}
This research is supported by the National Science and Technology Major Project (2023ZD0121102), National Natural Science Foundation of China (92270114, 62302321),
the Dutch Research Council (NWO), under project numbers 024.004.022, NWA.1389.20.\-183, and KICH3.LTP.20.006,
    and the European Union's Horizon Europe program under grant agreement No 101070212.

\balance
\bibliographystyle{ACM-Reference-Format}
\bibliography{references}

\clearpage

\end{document}